\documentclass[aps,prl,twocolumn,superscriptaddress]{revtex4}
\usepackage{graphicx}
\usepackage{bm}
\usepackage{color,soul}

\begin{document}

\title{$\omega / T $ scaling and magnetic quantum criticality in BaFe$_2$(As$_{0.7}$P$_{0.3}$)$_2$}

\author{Ding Hu}
\affiliation{Department of Physics and Astronomy,
Rice Center for Quantum Materials,
Rice University, Houston, Texas 77005-1827, USA}
\author{Haoyu Hu}
\affiliation{Department of Physics and Astronomy,
Rice Center for Quantum Materials,
Rice University, Houston, Texas 77005-1827, USA}
\author{Wenliang Zhang}
\affiliation{Beijing National Laboratory for Condensed Matter Physics, Institute of Physics, Chinese Academy of Sciences, Beijing 100190, China}
\affiliation{School of Physical Sciences, University of Chinese Academy of Sciences, Beijing 100190, China}
\author{Yuan Wei}
\affiliation{Beijing National Laboratory for Condensed Matter Physics, Institute of Physics, Chinese Academy of Sciences, Beijing 100190, China}
\affiliation{School of Physical Sciences, University of Chinese Academy of Sciences, Beijing 100190, China}
\author{Shiliang Li}
\affiliation{Beijing National Laboratory for Condensed Matter Physics, Institute of Physics, Chinese Academy of Sciences, Beijing 100190, China}
\affiliation{School of Physical Sciences, University of Chinese Academy of Sciences, Beijing 100190, China}
\affiliation{Songshan Lake Materials Laboratory , Dongguan, Guangdong 523808, China}
\author{Yanhong Gu}
\affiliation{Beijing National Laboratory for Condensed Matter Physics, Institute of Physics, Chinese Academy of Sciences, Beijing 100190, China}
\affiliation{School of Physical Sciences, University of Chinese Academy of Sciences, Beijing 100190, China}
\author{Xiaoyan Ma}
\affiliation{Beijing National Laboratory for Condensed Matter Physics, Institute of Physics, Chinese Academy of Sciences, Beijing 100190, China}
\affiliation{School of Physical Sciences, University of Chinese Academy of Sciences, Beijing 100190, China}
\author{Douglas L. Abernathy}
\affiliation{Neutron Scattering Division, Oak Ridge National Laboratory, Oak Ridge, TN 37831, USA.}
\author{Songxue Chi}
\affiliation{Neutron Scattering Division, Oak Ridge National Laboratory, Oak Ridge, TN 37831, USA.}
\author{Travis J. Williams}
\affiliation{Neutron Scattering Division, Oak Ridge National Laboratory, Oak Ridge, TN 37831, USA.}
\author{Yu Li}
\affiliation{Department of Physics and Astronomy,
Rice Center for Quantum Materials,
Rice University, Houston, Texas 77005-1827, USA}
\author{Qimiao Si}
\affiliation{Department of Physics and Astronomy,
Rice Center for Quantum Materials,
Rice University, Houston, Texas 77005-1827, USA}
\author{Pengcheng Dai}
\affiliation{Department of Physics and Astronomy,
Rice Center for Quantum Materials,
Rice University, Houston, Texas 77005-1827, USA}
\affiliation{Center for Advanced Quantum Studies and Department of Physics, Beijing Normal University, Beijing 100875, China}

\begin{abstract}
We used transport and inelastic neutron scattering to study
the optimally phosphorus-doped BaFe$_2$(As$_{0.7}$P$_{0.3}$)$_2$ superconductor ($T_c = 30$ K).
In the normal state, we find that the previously reported linear temperature dependence of the resistivity below room temperature extends to $\sim$ 500 K.
Our analysis of the temperature and energy ($E=\hbar\omega$) dependence of spin dynamical susceptibility
at the antiferromagnetic (AF) ordering wave vector $\chi^{\prime\prime}({\bf Q}_{\rm AF},\omega)$
reveal an $\omega / T$ scaling within $1.1<E/k_BT<110$. These results suggest that the linear temperature dependence of the
resistivity is due to the presence of a magnetic quantum critical point
in the cleanest iron pnictides near optimal superconductivity.
Moreover, the results reconcile the strange-metal temperature dependences
with the weakly first-order nature of the quantum transition out of the AF and nematic orders.
\end{abstract}


\maketitle

One of the hallmarks of unconventional superconductivity in copper oxide
superconductors is the linear temperature
dependence of the resistivity in the normal state below about 1000 K \cite{palee}.
First discovered in La$_{2-x}$Sr$_x$CuO$_4$ and YBa$_2$Cu$_3$O$_7$
nearly optimal superconductivity \cite{Cava1987,Gurvitch,Takagi,Ando,Louis2010,Hussey},
the linear temperature dependence
of the resistivity is incompatible with Landau's Fermi-liquid theory of metals, where
temperature dependence of the resistivity is expected to be quadratic ($T^2$) \cite{Hilbert2007},
and suggests
the presence of a quantum critical point (QCP) responsible for the breakdown
of Fermi-liquid behavior and
development of strange-metal properties \cite{Sachdev1995,Qimiao2001,Fradkin}.

In the case of iron pnictide superconductors \cite{Stewart2001,dai,Qimiao2016},
a linear temperature dependence of the resistivity has been found in different families of materials near optimal superconductivity, suggesting the presence of a QCP \cite{Chu2009,Zhou2013,Zhaoyu2016,Hosoi2016,Kuo2016,Fisher2012}. In particular, experimental evidence for a QCP in phosphorus-doped BaFe$_2$(As$_{1-x}$P$_x$)$_2$, envisioned in early theoretical studies \cite{Dai09}, has been mounting \cite{SJiang,Shibauchi2014}. This includes the linear temperature dependence of the resistivity \cite{kasa10}, a peak in the effective electron mass \cite{Shishido2010}, magnetic penetration depth \cite{Hashimoto}, heat capacity \cite{Walmsley} and nuclear magnetic resonance (NMR) \cite{Nakai}. The phosphorus doping does not involve the Fe-sites. As a result,  this series is especially clean, as demonstrated by the relatively small residual resistivity and the
observation of quantum oscillations \cite{Shishido2010}. Because the minimal disorder will allow for a clear understanding of the implications of the inelastic neutron scattering measurements (see below), here we focus on BaFe$_2$(As$_{1-x}$P$_x$)$_2$ near its optimal superconductivity.

In the undoped state, BaFe$_2$As$_2$ exhibits a tetragonal-to-orthorhombic
structural transition at $T_s$, where a nematic phase with in-plane electronic anisotropy is established \cite{CFang08,CXu,Dai09,Fernandes},
followed by a collinear antiferromagnetic (AF) order below $T_N \approx 140$ K ($\leq T_s$) at ${\bf Q}_{\rm AF}\approx (0.5,0.5)$
[Figs. 1(a), 1(b)] \cite{Stewart2001,dai}. When phosphorus is doped into BaFe$_2$As$_2$ to form BaFe$_2$(As$_{1-x}$P$_{x}$)$_2$ \cite{SJiang},
a QCP is found in BaFe$_2$(As$_{0.7}$P$_{0.3}$)$_2$ near optimal superconductivity with suppressed orthorhombic lattice
distortion and static AF order [Fig. 1(c)].
Increasing P-doping in BaFe$_2$(As$_{1-x}$P$_{x}$)$_2$ suppresses
both $T_s$ and $T_N$, which are associated nematic and AF phase transitions, respectively.
If both magnetic and nematic QCPs occur, one would expect to find gradually suppressed second order structural and magnetic phase transitions
with increasing P-doping in BaFe$_2$(As$_{1-x}$P$_{x}$)$_2$.
However, systematic neutron diffraction and NMR experiments on powder and single crystal  samples of BaFe$_2$(As$_{1-x}$P$_{x}$)$_2$ reveal that the structural and AF phase transitions are coupled at all $x$, and AF phase transition
around $x\approx 0.3$ become weakly first order near optimal superconductivity, thus suggesting an
avoided magnetic QCP \cite{Allred2014,DHu2015,Dioguardi2016}.

\begin{figure}
\includegraphics[scale=.8]{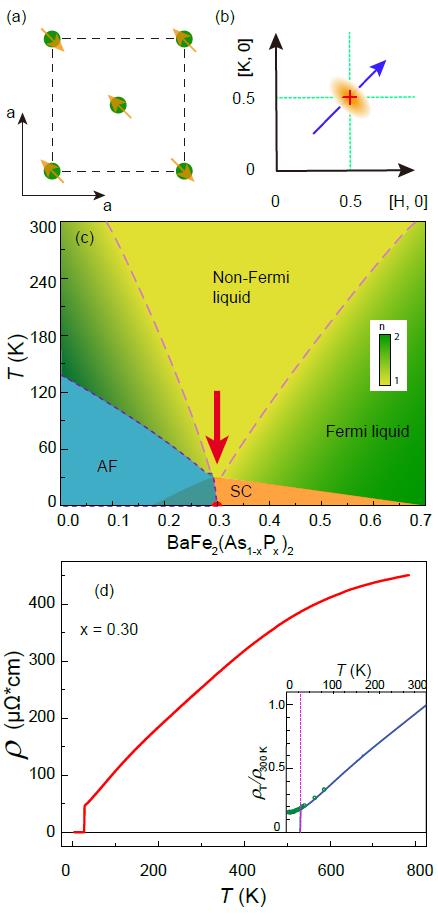}
\caption{(a) The AF structured Fe layer of
BaFe$_2$(As$_{1-x}$P$_x$)$_2$ in tetragonal notation, yellow arrows indicate the spin direction below $T_N$. (b) Position and shape of spin fluctuations in BaFe$_2$(As$_{0.7}$P$_{0.3}$)$_2$ in reciprocal space below 100 meV. Cross indicates the in-plane AF order position, blue arrow is the constant-energy scans direction measured on HB-3. (c) Schematic  BaFe$_2$(As$_{1-x}$P$_x$)$_2$ electronic phase diagram. Red arrow indicates the location of BaFe$_2$(As$_{0.7}$P$_{0.3}$)$_2$. (d) Resistivity ranges from 10 K to 790 K. Inset shows the resistivity measured under high magnetic field on similar concentration compound from \cite{fisher2014}. Pink dotted line shows the cutoff temperature from T-linear resistivity.
}
\label{fig1}
\end{figure}

Although the AF phase transition
in BaFe$_2$(As$_{1-x}$P$_{x}$)$_2$ is weakly first order near optimal
superconductivity \cite{Allred2014,DHu2015,Dioguardi2016}, the small ordered moment
and low ordering temperature do not exclude the possibility
of an extended
 temperature and energy range where
 quantum criticality underlies the linear resistivity.
This is consistent with the fact that when superconductivity
in BaFe$_2$(As$_{0.7}$P$_{0.3}$)$_2$ is suppressed by
a magnetic field, the temperature dependence of the resistivity
is quadratic
below the zero field $T_c=30$ K,
deviating from the linear temperature dependence above 30 K [see inset of Fig. 1(d)] \cite{fisher2014}.
Figure 1(d) shows
our measurement of
the resistivity
for BaFe$_2$(As$_{0.7}$P$_{0.3}$)$_2$ from
2 K to 790 K. In addition to confirming the linear temperature dependence of the
resistivity from 30 K to 300 K \cite{kasa10}, the data reveal
that it extends all the way up to $\sim$ 500 K, above which a
clear deviation from
the linear behavior is seen; this exemplifies the extended temperature range for the
strange-metal behavior.
The nematic QCP \cite{Kuo2016} alone is unlikely to be responsible for
the observed linear temperature dependent resistivity \cite{Shibauchi2014}, given that fluctuations
at small wavevectors are inefficient in degrading an electrical current.
If linear temperature
dependence of the resistivity in BaFe$_2$(As$_{0.7}$P$_{0.3}$)$_2$ in Fig. 1(d)
is
associated
with a magnetic quantum critical fluctuations, one would expect that spin dynamics
at ${\bf Q}_{\rm AF}$ to follow $\omega /T$ scaling within a finite energy ($E=\hbar\omega$,
where $\omega$ is frequency) and temperature range \cite{stockert2011,dai2005,dai2006,Kim2015}.

\begin{figure}
\includegraphics[scale=0.7]{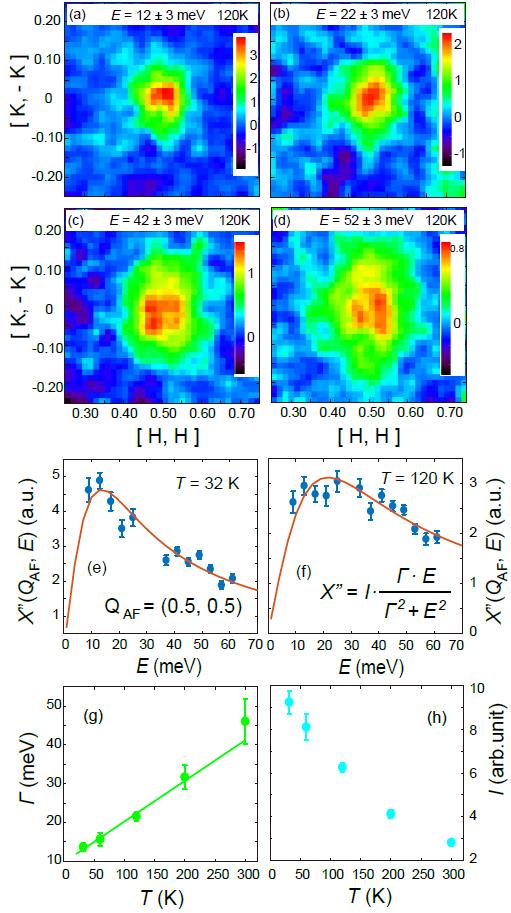}
\caption{Images of two-dimensional spin fluctuations with energy transfers (a) $E=12\pm 3$, (b) $22\pm 3$, (c) $42\pm 3$, and (d) $52\pm 3$ meV, obtained on ARCS after the background subtraction and Bose factor correction. The color bars in (a)-(d) indicate intensity in absolute units of mbarn sr$^{-1}$meV$^{-1}$f.u.$^{-1}$. Energy dependence of the imaginary part of the dynamic susceptibility at ${\bf Q}_{\rm AF}$  are obtained by fitting these ARCS data collected at (e) 32 K, and (f) 120 K with two-dimensional Gaussian function. Red lines are fitting results as discussed in the text. (g) and (h) show the temperature dependence of $\Gamma$ and I. Green line in (g) is the fitting result of linear function.
}
\label{fig2}
\end{figure}

To test this hypothesis, we use inelastic neutron
scattering to study BaFe$_2$(As$_{0.7}$P$_{0.3}$)$_2$
($T_c \approx 30$ K), focusing on temperature and energy dependence of spin fluctuations near
${\bf Q}_{\rm AF}$. In previous inelastic neutron scattering experiments
on BaFe$_2$(As$_{0.7}$P$_{0.3}$)$_2$ \cite{CHLee13,ding2016,ding2017},
a $c$-axis dispersive neutron spin resonance coupled to superconductivity
has been identified. By using neutron triple-axis and time-of-flight spectroscopy, we
find that energy and temperature dependence of the
imaginary part of dynamic susceptibility at ${\bf Q}_{\rm AF}$
$\chi^{\prime\prime}({\bf Q}_{\rm AF},\omega)$, which is related to magnetic
scattering $S({\bf Q}_{\rm AF},\omega)$ via
$\chi^{\prime\prime}({\bf Q}_{\rm AF},\omega)\propto (1-e^{-\hbar\omega/k_BT})S({\bf Q}_{\rm AF},\omega)$, follows the $\omega /T$ scaling for
$9<E<61$ meV and 5 K$<T< 200$ K. For energies less than about $E=10$ meV,
the $\omega /T$ scaling fails to describe the data. These results suggest that the
observed linear temperature dependence of the resistivity in Fig. 1(d)
may arise from magnetic quantum critical fluctuations important for controlling
the transport and nematic properties of BaFe$_2$(As$_{0.7}$P$_{0.3}$)$_2$.

\begin{figure}
\includegraphics[scale=0.8]{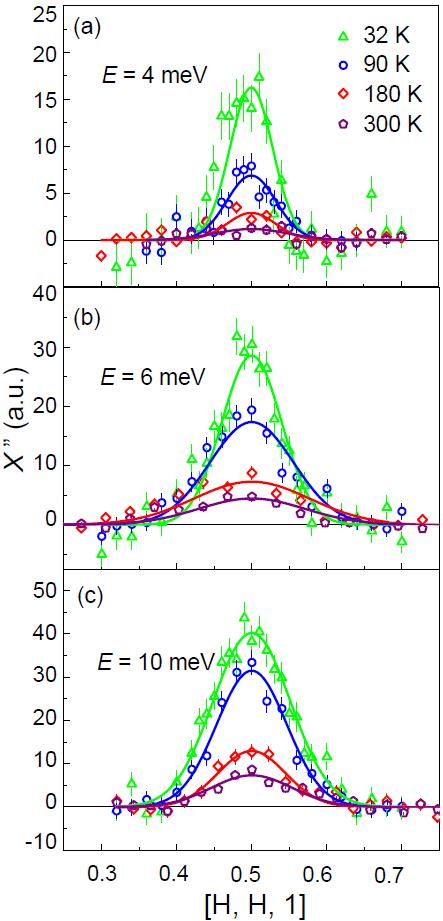}
\caption{Low energy spin fluctuations in BaFe$_2$(As$_{0.3}$P$_{0.7}$)$_2$ at different temperatures. Constant energy scans along the longitudinal direction through ${\bf Q}_{\rm AF}$ with energy transfers of (a) 4, (b) 6, and (c) 10 meV. Backgrounds and Bose factor are corrected for all data. The error bars indicate the statistical errors of one standard deviation.
}
\label{fig3}
\end{figure}

We used the standard four-probe method to measure the resistivity of
BaFe$_2$(As$_{0.3}$P$_{0.7}$)$_2$
 from 10 K to 790 K in a Janis 4 K closed cycle refrigerator with high temperature capability.
As we can see from Fig. 1(d),
 the remarkable linear temperature dependence of the resistivity
extends
up to 500 K.
Combined with
previous transport measurements below 30 K when superconductivity is suppressed by a high magnetic
field [see inset of Fig1(d)] \cite{fisher2014}, we find
that the temperature range
for linear resistivity
 is from 30 K to 500 K.

Our inelastic neutron scattering experiments were carried out on the Wide Angular-Range Chopper Spectrometer (ARCS) at the Spallation Neutron Source and HB-3 triple axis spectrometer at the High-Flux Isotope Reactor, both at Oak Ridge National Laboratory. For the time-of-flight measurement on ARCS, we used $E_i = 80$ meV with $k_i$ parallel to the $c$ axis.
Total mass 17 g high-quality BaFe$_2$(As$_{0.3}$P$_{0.7}$)$_2$ single crystals were co-aligned in the $[H,H,L]$ scattering plane with an in-plane mosaic $<$ 5.5$^\circ$ \cite{ding2017}.
We
define $Q = (H, K, L) = \frac{2\pi}{a}H\hat{i} + \frac{2\pi}{a}K\hat{j} +
\frac{2\pi}{a}L\hat{k}$, where the tetragonal lattice constants are $a = b \approx 3.96 \AA$ and $c \approx 12.87 \AA$.

Figures 2(a)-(d) show images of spin excitations $S({\bf Q},\omega)$ at $E=12\pm 3$, $22\pm 3$,
$42\pm 3$, $52\pm 3$ meV, respectively, at $T=120$ K. Consistent
with earlier work \cite{ding2016}, spin excitations form transversely elongated ellipses
centered around ${\bf Q}_{\rm AF}$
that disperse outward with increasing energy. We convert $S({\bf Q},\omega)$
to $\chi^{\prime\prime}({\bf Q},\omega)$ and fit
the in-plane dynamical susceptibility with two-dimensional Gaussian function to get the absolute
intensity of $\chi^{\prime\prime}({\bf Q},\omega)$  at ${\bf Q}={\bf Q}_{\rm AF}=(0.5,0.5)$.
Figures 2(e) and 2(f) show energy dependence of  $\chi^{\prime\prime}({\bf Q}_{\rm AF},\omega)$
at $T=32$ and 120 K, respectively. The solid lines in the figures are fits to
the data with typical paramagnetic relaxational form
$\chi^{\prime\prime}({\bf Q}_{\rm AF},\omega)=I\frac{\Gamma * E}{\Gamma ^2 + E^2}$,
where $\Gamma$ is related to full width at half maximum of the excitations,
corresponding to the relaxation lifetime of the excitations, and $I$ is the peak intensity of the excitations.
Figures 2(g) and 2(h) show temperature dependence of $\Gamma$ and $I$, respectively.
While $\Gamma$ increases approximately linearly with increasing temperature, $I$
decreases with increasing temperature.

\begin{figure}
\includegraphics[scale=0.3]{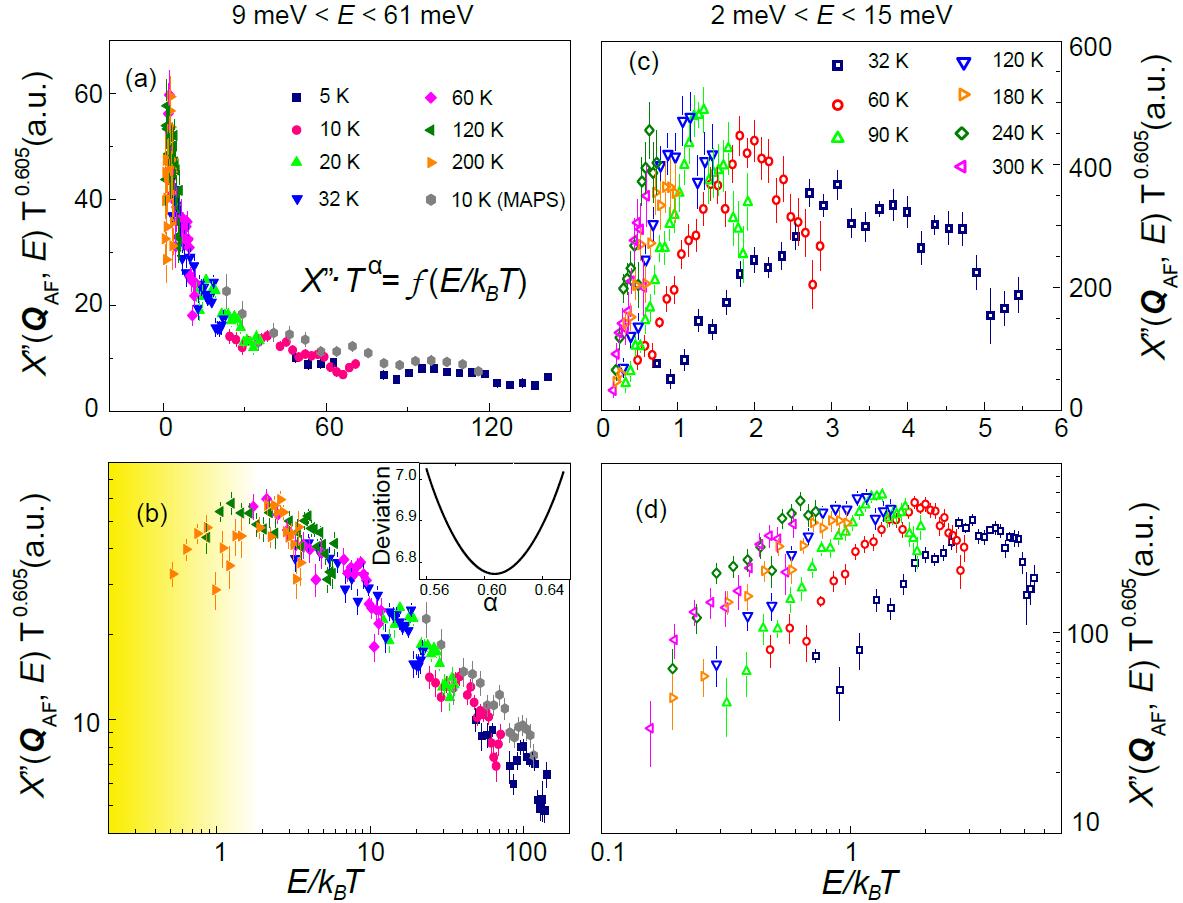}
\caption{The $\omega/T$ scaling plot of $\chi^{\prime\prime}({\bf Q}_{\rm AF},\omega)$ for data obtained at ARCS and HB-3. (a) and (b) show the data collected on ARCS with linear and logarithmic scales, respectively. Data below 20 meV for 5, 10 and 20 K have not been added in figure to eliminate the influence of superconductivity. The inset shows the quality of the data collapse for different values of the critical exponent $\alpha$. Best scaling is found for $\alpha$ = 0.605. Shaded area in (b) indicates the region where $\omega/T$ scaling is not obeyed. Gray points are the data from 20 to 100 meV collected on MAPS at 10 K with $E_i = 250 $ meV. Plots (c) and (d) are data from HB-3.}
\label{fig4}
\end{figure}

To test if the measured imaginary part of the dynamic susceptibility at ${\bf Q}_{\rm AF}$ follows the $\omega / T$ scaling expected for magnetic quantum critical fluctuations, we consider $\chi^{\prime\prime}({\bf Q}_{\rm AF},\omega)T^{\alpha}=F({\bf Q}_{\rm AF}, \omega/T)$, where the scaling exponent $\alpha$ and the scaling function $F ({\bf Q}_{\rm AF}, \omega / T)$ are determined through the best-observed collapse of the data onto one universal curve \cite{dai2006}. We exclude the data under $E=20$ meV below 30 K to eliminate the impact of superconductivity on $\chi^{\prime\prime}({\bf Q}_{\rm AF},\omega)$ due to the appearance of the neutron spin resonance in superconducting state \cite{CHLee13,ding2016}. By fitting the data with this function,
we find $\alpha = 0.605$ independent of the functional form of $F({\bf Q}_{\rm AF},\omega / T)$ [Fig. 4 (a),(b)]. For $E/T$ from 1.1 to about 110,
the data collapse into a single curve. From neutron time-of-flight measurements
on BaFe$_2$(As$_{0.7}$P$_{0.3}$)$_2$, we know that spin excitations disperse transversely away from ${\bf Q}_{\rm AF}$ for energies above 100 meV. Figures 4(a) and 4(b) suggest that spin excitations up to 100 meV follow $\omega / T$ scaling. However, low-energy spin excitations ($E/T<1.1$) seem to deviate from the scaling curve, consistent with transport measurements where resistivity deviates from linear temperature dependence below 30 K.

To further explore if the low-energy spin fluctuations follow $\omega / T$ scaling,
we carried out inelastic neutron scattering measurements using HB-3 triple axis spectrometer. Figure 3 shows $\chi^{\prime\prime}({\bf Q},\omega)$ for
$E = 4, 6, 10 $ meV at various temperatures. The spin fluctuations in these energies are fitted to a Gaussian function and give the dynamical susceptibility at ${\bf Q}_{\rm AF}$. The spin fluctuations at $E = 4, 6, 10$ meV are dramatically suppressed with increasing temperature, but still present even at 300 K [Figs. 3(a)-(c)].
By combining results from constant-energy and constant-${\bf Q}$ scans, we deduce the data
using the same parameters as in Figs. 4(a) and 4(b) and the outcome is shown in
Figs. 4(c) and 4(d) with $\alpha = 0.605$. Clearly, the $\omega / T$ scaling fails
for low-energy spin fluctuations with energy approximately below 10 meV.

We have shown that the energy and temperature dependent of
$\chi^{\prime\prime}({\bf Q}_{\rm AF},\omega)$
 in BaFe$_2$(As$_{0.7}$P$_{0.3}$)$_2$ follows $\omega / T$ scaling within
an extended temperature and energy range.
To appreciate the importance of this material being clean, we recall that, in
the case of doped
heavy-Fermion materials
such as UCu$_{5-x}$Pd$_x$, the observed
$\omega / T$ and associated non-Fermi liquid behavior such as linear temperature
dependence of the resistivity have been attributed to either
quantum criticality of
long-range magnetic order or localized moments in a disordered setting
\cite{Hilbert2007,MacL2004,Miranda1996,Bernal1995,Aronson95}.
In Cu-doped Ba(Fe$_{1-x}$Cu$_x$)$_2$As$_2$, the AF order found in parent compound
broadens with increasing Cu-substitution,
suggesting the occurrence of a spin-glass state \cite{WYWang2017}.
The $\omega / T$ scaling found in Ba(Fe$_{1-x}$Cu$_x$)$_2$As$_2$
can be understood by the theoretical models for the AF QCP or spin-glass QCP \cite{Kim2015}.
Recently, the existence of
an
AF QCP was suggested in Ba(Fe$_{0.97}$Cr$_{0.03}$)$_2$(As$_{1-x}P_x$)$_2$ at $x = 0.42$
based on
$\omega / T$ scaling \cite{shiliang2018}.
However, Cr doping
in BaFe$_2$(As$_{1-x}$P$_{x}$)$_2$
induces significant impurity scattering at the Fe position.
It is therefore unclear if
the observed non-Fermi liquid behavior and $\omega / T$ scaling at $x$ = 0.42,
which is clearly different from the
behavior at $x=0.3$
without Cr doping,
are due to impurity
scatterings from disordered moments of Cr spins.
By contrast, BaFe$_2$(As$_{1-x}$P$_{x}$)$_2$ has
negligible disorder \cite{Shishido2010}
given that the P-substitution
does not go into the Fe-plane,
and this is
consistent with the instrumental resolution limited
magnetic Bragg peaks up to $x = 0.3$ \cite{DHu2015}.
 Therefore, the observed $\omega / T$ scaling
we have observed over an extended
 temperature and energy range in
the
 $x=0.3$ compound
must be attributed
to quantum spin fluctuations related
to the vanishing weakly first order AF phase transition.
The violation of the $\omega/T$ scaling observed
below 10 meV adds further support to this interpretation.

Quantum criticality associated with a weakly first-order but concurrent AF and nematic quantum phase transitions were anticipated theoretically for the P-substituted iron arsenides \cite{Dai09,Wu16}. P-for-As substitution effectively decreases the strength of electron correlations by enhancing the kinetic energy, thereby reducing the magnetic and the concomitant nematic order. This effect can be analyzed by a field theory containing both the magnetic and nematic degrees of freedom. Expressed in terms of
${\vec m}_{A/B}$, the staggered magnetization of the spins of the $3d$-electrons on the $A$ and $B$ sublattices of the Fe-square lattice, the $(1/2,1/2)$/$(1/2,-1/2)$ AF degree of freedom is  $({\vec m}_{A} \pm {\vec m}_{B})/2$ while the
nematic one is ${\vec m}_{A} \cdot {\vec m}_{B}$. The effective theory contains an interaction term,
$- u_I
 ( {\vec m} _A
\cdot
{\vec m} _B
)
^2$, with
the coupling $u_I>0$.
In the renormalization group sense,
this interaction term is relevant with respect to an underlying magnetic
QCP, but only marginally so. Consequently, it drives both the AF and
nematic quantum phase transitions to be concurrent and weakly first order, resulting in
a large dynamical range of quantum criticality \cite{Dai09}. The same conclusion follows from
a large-$N$ saddle-point calculation \cite{Wu16}.

In summary, we have systematically measured the temperature and energy evolution of spin fluctuations in the optimal doped BaFe$_2$(As$_{0.7}$P$_{0.3}$)$_2$.
We find evidence for the $\omega / T$ scaling in spin fluctuations
over
extended
temperature and energy range,
consistent with linear temperature dependence of the resistivity. These results provided
strong evidence that the linear temperature dependence of resistivity arises
BaFe$_2$(As$_{0.7}$P$_{0.3}$)$_2$,
which is located near optimal superconductivity,
 from
magnetic quantum criticality.
 Therefore, the presence of a magnetic QCP may ultimately be responsible for the anomalous transport and magnetic properties of
the iron pnictides and strongly influence their superconductivity.

The neutron scattering work at Rice is supported by the
U.S. NSF-DMR-1700081 (P.D.). A part of the material synthesis work at Rice is
supported by the Robert A. Welch Foundation Grant No. C-1839 (P.D.).
The theoretical work at Rice is supported by the U.S. Department of Energy, Office of Science,
Basic Energy Sciences, under Award No. DESC0018197, and the Robert A.Welch Foundation Grant No. C-1411.
Q.S. acknowledges  the hospitality and the support by a Ulam Scholarship
of the Center for Nonlinear Studies at Los Alamos National Laboratory.
The transport work at IOP is supported by the Ministry of Science and Technology of China (No. 2017YFA0302903, No. 2016YFA0300502) and the National Natural Science Foundation of China (No. 11674406).
This research used resources at the High Flux Isotope Reactor and Spallation Neutron Source, DOE Office of Science User Facilities operated by the Oak Ridge National Laboratory.


\begin{thebibliography}{}

\bibitem{palee} P. A. Lee, N. Nagaosa, and X. G. Wen, Rev. Mod. Phys. {\bf 78}, 1785 (2006).

\bibitem{Cava1987} R. J. Cava, R. B. van Dover, B. Batlogg, and E. A. Rietman, Phys. Rev. Lett. \textbf{58}, 408 (1987).

\bibitem{Gurvitch} M. Gurvitch and A. T. Fiory, Phys. Rev. Lett. {\bf 59}, 1377 (1987).


\bibitem{Takagi} H. Takagi, B. Batlogg, H. L. Kao, J. Kwo, R. J. Cava, J. J. Krajewski, and W. F. Peck, Jr.,
Phys. Rev. Lett. {\bf 69}, 2975 (1992).


\bibitem{Ando} Y. Ando, S. Komiya, K. Segawa, S. Ono, Y. Kurita, Phys. Rev. Lett. {\bf 93}, 267001 (2004).

\bibitem{Louis2010} Louis Taillefer, Annu. Rev. Condens. Matter Phys. \textbf{1}, 51 (2010).

\bibitem{Hussey} N. E. Hussey, R. A. Cooper, Xiaofeng Xu, Y. Wang, I. Mouzopoulou, B. Vignolle, C. Proust, Phil. Trans. R. Soc. A {\bf 369}, 1626 (2011).

\bibitem{Hilbert2007} Hilbert v. L$\ddot{o}$hneysen, Achim Rosch, Matthias Vojta,
Peter W$\ddot{o}$lfle, Rev. Mod. Phys. \textbf{79}, 1015 (2007).

\bibitem{Sachdev1995} Subir Sachdev, Phys. Status Solidi B {\bf 247}, 537 (2010).

\bibitem{Qimiao2001} Q. Si, S. Rabello, K. Ingersent and J. L. Smith, Nature (London) \textbf{413}, 804 (2001).

\bibitem{Fradkin} E. Fradkin, S. A. Kivelson, and J. M. Tranquada, Rev. Mod. Phys. {\bf 87}, 457 (2015).

\bibitem{Stewart2001} G. R. Stewart, Rev. Mod. Phys. \textbf{73}, 797 (2001).

\bibitem{dai} Pengcheng Dai, Rev. Mod. Phys. {\bf 87}, 855 (2015).

\bibitem{Qimiao2016} Qimiao Si, Rong Yu and Elihu Abrahams, Nature Rev. Mater. \textbf{1}, 16017 (2016).

\bibitem{Chu2009} Jiun-Haw Chu, James G. Analytis, Chris Kucharczyk, and Ian R. Fisher, Phys. Rev. B \textbf{79}, 014506 (2009).

\bibitem{Zhou2013} R. Zhou, Z. Li, J. Yang, D.L. Sun, C.T. Lin and Guo-qing Zheng, Nat. Commun.  \textbf{4}, 2265 (2013).

\bibitem{Zhaoyu2016} Zhaoyu Liu, Yanhong Gu, Wei Zhang, Dongliang Gong, Wenliang Zhang, Tao Xie, Xingye Lu, Xiaoyan Ma, Xiaotian Zhang, Rui Zhang, Jun Zhu, Cong Ren, Lei Shan, Xianggang Qiu, Pengcheng Dai, Yi-feng Yang, Huiqian Luo, and Shiliang Li, Phys. Rev. Lett. \textbf{117}, 157002 (2016).

\bibitem{Hosoi2016} Suguru Hosoia, Kohei Matsuuraa, Kousuke Ishidaa, Hao Wanga, Yuta Mizukamia, Tatsuya Watashigeb, Shigeru Kasaharab, Yuji Matsudab, and Takasada Shibauchia, PNAS, \textbf{113}, 8139-8143 (2016).

\bibitem{Kuo2016} Hsueh-Hui Kuo, Jiun-Haw Chu, Johanna C. Palmstrom, Steven A. Kivelson, Ian R. Fisher, Science  \textbf{4}, 958 (2016).

\bibitem{Fisher2012} Jiun-Haw Chu, Hsueh-Hui Kuo, James G. Analytis, Ian R. Fisher, Science  \textbf{337}, 710 (2012).


\bibitem{Dai09} J. Dai, Q. Si, J.-X. Zhu, and E. Abrahams, PNAS {\bf 106}, 4118 (2009).



\bibitem{SJiang} S. Jiang, C. Wang, Z. Ren, Y. Luo, G. Cao, and Z.-A. Xu, J. Phys. Condens. Matter {\bf 21}, 382203 (2009).

\bibitem{Shibauchi2014} T. Shibauchi, A. Carrington, and Y. Matsuda, Annu. Rev. Condens.Matter Phys. \textbf{5}, 113 (2014).

\bibitem{kasa10} S. Kasahara, T. Shibauchi, K. Hashimoto, K. Ikada, S. Tonegawa, R. Okazaki, H. Shishido, H. Ikeda, H. Takeya, K. Hirata, T. Terashima, and Y. Matsuda Phys. Rev. B \textbf{81}, 184519 (2010).

\bibitem{Shishido2010} H. Shishido, A. F. Bangura, A. I. Coldea, S. Tonegawa, K. Hashimoto, S. Kasahara, P. M. C. Rourke, H. Ikeda, T. Terashima, R. Settai, Y. $\bar{O}$ nuki, D. Vignolles, C. Proust, B. Vignolle, A. McCollam, Y. Matsuda, T. Shibauchi, and A. Carrington, Phys. Rev. Lett. \textbf{104}, 057008 (2010).

\bibitem{Hashimoto} K. Hashimoto {\it et al.}, Science {\bf 336}, 1554 (2012).

\bibitem{Walmsley} P. Walmsley {\it et al.}, Phys. Rev. Lett. {\bf 110}, 257002 (2013).

\bibitem{Nakai} Y. Nakai, T. Iye, S. Kitagawa, K. Ishida, H. Ikeda,
S. Kasahara, H. Shishido, T. Shibauchi, Y. Matsuda, and
T. Terashima, Phys. Rev. Lett. {\bf 105}, 107003 (2010).

\bibitem{CFang08} C. Fang, H. Yao, W. Tsai, J. P. Hu, and S. A. Kivelson, Phys. Rev.
B {\bf 77}, 224509 (2008).

\bibitem{CXu} C. Xu, M. M$\rm \ddot{u}$ller, and S. Sachdev, Phys. Rev. B {\bf 78}, 020501
(2008).

\bibitem{Fernandes} R. M. Fernandes, A. V. Chubukov, and J. Schmalian, Nat. Phys. {\bf 10}, 97 (2014).

\bibitem{Allred2014} J. M. Allred, K. M. Taddei, D. E. Bugaris, S. Avci, D. Y. Chung, H. Claus, C. dela Cruz, M. G. Kanatzidis, S. Rosenkranz, R. Osborn, and O. Chmaissem, Phys. Rev. B \textbf{90}, 104513 (2014).

\bibitem{DHu2015} D. Hu, X. Y. Lu, W. L. Zhang, H. Q. Luo, S. L. Li, P. P. Wang, G. F. Chen, F. Han, S. R. Banjara, A. Sapkota, A. Kreyssig, A. I. Goldman, Z. Yamani, Ch. Niedermayer, M. Skoulatos, R. Georgii, T. Keller, P. S. Wang, W. Q. Yu, and P. C. Dai, Phys. Rev. Lett. {\bf 114},157002(2015).

\bibitem{Dioguardi2016} A. P. Dioguardi, T. Kissikov, C. H. Lin, K. R. Shirer, M. M. Lawson, H.-J. Grafe, J.-H. Chu, I. R. Fisher, R. M. Fernandes, and N. J. Curro, Phys. Rev. Lett. {\bf 116}, 107202 (2016).

\bibitem{fisher2014} James G. Analytis, H-H. Kuo, Ross D. McDonald, MarkWartenbe, P. M. C. Rourke, N. E. Hussey and I. R. Fisher, Nature Phys.  \textbf{10}, 194-197 (2014).

\bibitem{stockert2011} O. Stockert and F. Steglich,  Annu. Rev. Condens.Matter Phys. \textbf{2}, 79 (2011).

\bibitem{dai2005} Stephen D. Wilson, Pengcheng Dai, D. T. Adroja, S.-H. Lee, J.-H. Chung, J.W. Lynn, N. P. Butch, and M. B. Maple, Phys. Rev. Lett. {\bf 94}, 056402 (2005).

\bibitem{dai2006} Stephen D. Wilson, Shiliang Li, Pengcheng Dai, Wei Bao, Jae-Ho Chung, H. J. Kang, Seung-Hun Lee, Seiki Komiya, Yoichi Ando, and Qimiao Si, Phys. Rev. B \textbf{74}, 144514 (2006).

\bibitem{Kim2015} M. G. Kim, M. Wang, G. S. Tucker, P. N. Valdivia, D. L. Abernathy, Songxue Chi, A. D. Christianson, A. A. Aczel, T. Hong, T. W. Heitmann, S. Ran, P. C. Canfield, E. D. Bourret-Courchesne, A. Kreyssig, D. H. Lee, A. I. Goldman, R. J. McQueeney, and R. J. Birgeneau, Phys. Rev. B {\bf 92}, 214404 (2015).

\bibitem{CHLee13} C. H. Lee, P. Steffens, N. Qureshi, M. Nakajima, K. Kihou, A. Iyo, H. Eisaki, and M. Braden, Phys. Rev. Lett. {\bf 111}, 167002 (2013).

\bibitem{ding2016} Ding Hu, Zhiping Yin, Wenliang Zhang, R. A. Ewings, Kazuhiko Ikeuchi, Mitsutaka Nakamura, Bertrand Roessli, Yuan Wei, Lingxiao Zhao, Genfu Chen, Shiliang Li, Huiqian Luo, Kristjan Haule, Gabriel Kotliar, and Pengcheng Dai, Phys. Rev. B \textbf{94}, 094504 (2016).

\bibitem{ding2017} Ding Hu, Wenliang Zhang, Yuan Wei, Bertrand Roessli, Markos Skoulatos, Louis Pierre Regnault, Genfu Chen, Yu Song, Huiqian Luo, Shiliang Li, and Pengcheng Dai, Phys. Rev. B \textbf{96}, 180503(R) (2017).



\bibitem{Cooper} R. A. Cooper, Y. Wang, B. Vignolle, O. J. Lipscombe, S. M. Hayden, Y. Tanabe, T. Adachi, Y. Koike, M. Nohara, H. Takagi, Cyril Proust, N. E. Hussey,  Science {\bf 323}, 5914, 603-607 (2009).

\bibitem{MacL2004} D. E. MacLaughlin, R. H. Heffner, O. O. Bernal, K. Ishida, J. E. Sonier, G. J. Nieuwenhuys, M. B. Maple, and G. R. Stewart, J. Phys.: Condens. Matter Matter {\bf 16}, S4479 (2004).

\bibitem{Miranda1996} E. Miranda, V. Dobrosavljevic, and G. Kotliar, J. Phys.: Condens. Matter {\bf 8}, 9871 (1996).

\bibitem{Bernal1995} O. O. Bernal, D. E. MacLaughlin, H. G. Lukefahr, and B. Andraka, Phys. Rev. Lett. {\bf 75}, 2023 (1995).

\bibitem{Aronson95} M. C. Aronson, R. Osborn, R. A. Robinson, J. W. Lynn, R. Chau, C. L. Seaman, and M. B. Maple, Phys. Rev. Lett. {\bf 75}, 725 (1995).

\bibitem{WYWang2017} W. Y. Wang, Y. Song, D. Hu, Y. Li, R. Zhang, L. W. Harriger, W. Tian, H. B. Cao, and P. C. Dai, Phys. Rev. B {\bf 96}, 161106(R) (2017).


\bibitem{shiliang2018} Wenliang Zhang, Yuan Wei, Tao Xie, Zhaoyu Liu, Dongliang Gong, Xiaoyan Ma, Ding Hu, Petr $\rm \check{C}$erm$\rm \acute{a}$k, Astrid Schneidewind, Gregory Tucker, Siqin Meng, Zita H$\rm \ddot{u}$esges, Zhilun Lu, Jianming Song, Wei Luo, Liangcai Xu, Zengwei Zhu, Xunqing Yin, Hai-Feng Li, Yi-feng Yang, Huiqian Luo, and Shiliang Li, Phys. Rev. Lett. (in press).

\bibitem{Wu16} J. Wu, Q. Si, and E. Abrahams,
Phys. Rev. B {\bf 93}, 104515 (2016).





\end{thebibliography}
\end{document}